\documentclass[prb,a4paper,twocolumn,floatfix]{revtex4}
\usepackage[T1]{fontenc}
\usepackage{times} 
\usepackage{graphicx}
\usepackage{bm}
\begin{document}



\title{Magnetic origin of chemical balance in alloyed Fe-Cr stainless steels: first-principles and Ising model study}

\author{E. Airiskallio}
\author{E. Nurmi}
\author{I. J. V\"ayrynen}
\author{K. Kokko} 
\email{kalevi.kokko@utu.fi}
\affiliation{Department of Physics and Astronomy, University of Turku, FI-20014
Turku, Finland}
\affiliation{Turku University Centre for Materials and Surfaces (MatSurf), Turku, Finland}
\author{M. Ropo}
\affiliation{Department of Information Technology, \AA bo Akademi, FI-20500 Turku, Finland}
\author{M. P. J. Punkkinen}
\affiliation{Department of Physics and Astronomy, University of Turku, FI-20014
Turku, Finland}
\affiliation{Applied Materials Physics, Department of Materials
Science and Engineering, Royal Institute of Technology,
SE-10044 Stockholm, Sweden}
\author{B. Johansson}
\affiliation{Applied Materials Physics, Department of Materials
Science and Engineering, Royal Institute of Technology, 
SE-10044 Stockholm, Sweden}
\affiliation{Department of Physics and Materials Science, Uppsala University,
SE-75121 Uppsala, Sweden} 
\author{L. Vitos}
\affiliation{Applied Materials Physics, Department of Materials
Science and Engineering, Royal Institute of Technology, 
SE-10044 Stockholm, Sweden}
\affiliation{Department of Physics and Materials Science, Uppsala University, Box 516, SE-75120 Uppsala, Sweden} 
\affiliation{Research Institute for Solid State Physics and Optics, Budapest
H-1525, P.O. Box 49, Hungary}

\date{11 January 2012}

\begin{abstract}
Iron-chromium forms the basis of most of the stainless steel grades in the markets. Recently new insights into the physical and chemical properties of Fe-Cr based alloys have been obtained. Some of the new results are quite unexpected and call for further investigations. The present study addresses the magnetic contribution in the atomic driving forces behind the chemical composition in Fe-Cr alloyed with Al, Ti, V, Mn, Co, Ni, and Mo. Using the {\em ab initio} exact muffin-tin orbitals method and an Ising-type spin model, it is found that the magnetic moment of the solute atom combined with the induced changes in the magnetic moments of the host atoms form the main framework in determining the mixing energy and chemical potentials of low-Cr Fe-Cr based alloys. The results obtained in the present work are related to tuning of the microstructure and corrosion protection of low-Cr steels.
\end{abstract}


\maketitle

\section{Introduction}

The superb properties of stainless steel, like  resistance to corrosion, high strength, ductility, low maintenance, and relatively low cost make it an ideal base material for a host of commercial applications. For example, stainless steels are used in cook-ware, cutlery, hardware, surgical instruments, major appliances, industrial equipment, as an automotive 
and aerospace structural alloy, and construction material in large buildings.

Stainless steel differs from corrodible carbon steel mainly by the amount of chromium in the bulk. Stainless steels have sufficient amount of chromium present so that a thin and transparent film of chromium oxide rapidly forms to the open surface of the metal which prevents further surface corrosion and blocks corrosion from spreading into the internal structure of the material. More importantly, this oxide layer quickly reforms when the surface is scratched. This phenomenon is called 'self-healing passivation' and is seen also in other metals, such as aluminum and titanium.

High oxidation-resistance in air at ambient temperature is normally achieved with chromium additions in steels. The observed steep increase of the corrosion resistance of the ferritic stainless steels starts when the Cr content in bulk reaches the level of about 10 at.\%.\cite{Wranglen1985} This bulk threshold of Cr correlates surprisingly well with the calculated reversal point of the magnitudes of the bulk and surface effective chemical potential differences. This reversal of the chemical potentials is the cause for the outburst of Cr on the otherwise pure Fe surface of Fe-Cr alloys.\cite{Ropo07} In addition to Cr many other elements are frequently used as minor alloying additions in stainless steels to further improve the properties of stainless steels grades. However their impact on the chemical potential of the host matrix is not well documented.

Alloying elements in steels can be divided into two main categories namely austenite and ferrite stabilizers. From the elements discussed in the present work  Cr, Ti, Mo, V, and Al belong to the ferrite stabilizers and Ni, Mn, and Co are austenite stabilizers. In addition to the effect on the austenite/ferrite stabilization, these elements affect many other properties of steels. In the following we list some typical applications of alloying. Aluminum is often used in high temperature corrosion resistant
materials\cite{case_1953,khanna_2002,khanna_2005} due to its ability to form a highly stable and protective oxide scale on the open surface when exposed to oxidizing environment. The physical properties and corrosion resistance of Fe-Cr-Al as a function of the chemical composition of the alloy have been studied quite extensively.\cite{tomaszewicz_1978,gordon_1979,abderrazik_1987,stott_1987,stott_1989,prescott_1992,devan_1993,gurrappa_2000,schwalm_2000,babu_2001,lee_2003,wright_2004,nychka_2005,rohr_2005,grabke_2006,asteman_2008,Niu08} Titanium is an element added to steels because it increases the strength and resistance to corrosion. Furthermore, Ti provides a desirable property to alloys: lightness. Its density is less than half that of steel, so a titanium-steel alloy weighs less than pure steel and is more durable and stronger. These physical and mechanical properties make titanium steels a possible material for example in  aircraft and spacecraft engineering, chemical industry, and medicine.\cite{cismaru1964,bennet1979,freiburg1991,choi2005,firstov2009} Vanadium is a common component, for instance, in tool steels. Vanadium addition to Fe-Cr increases the hardness, tensile strength, wear resistance, and impact resistance.\cite{Unkic2003,Tjong1991} However, little has been published on the corrosion properties of the Fe-Cr-V alloys and on the role of vanadium in the passivation process.\cite{Brookes1990,Ras02,Chaliampalias2009} Molybdenum is used to increase the strength to weight ratio and weldability.\cite{michel2011} 

Nickel is the most frequent alloying element to stabilize Fe-Cr alloys in the austenite structure. This crystal structure makes such steels paramagnetic and less brittle at ambient conditions. Also manganese has been used in many stainless steel compositions. Manganese preserves the austenite structure in steels as does nickel, but at a lower cost. Austenite grades are perhaps the most widely used stainless steels in modern applications.\cite{garner1987,cismaru1964} Cobalt is used to increase heat and wear resistance (improve anti-galling property) and to produce excellent magnetic properties and good ductility.\cite{tjong1990}

In this Report, we investigate the role of minor alloying elements (Al, Ti, V, Mn, Co, Ni, Mo) on the relative energetics of Fe and Cr and its possible consequences to surface properties. The basic data is obtained by an {\em ab initio} electronic structure method and in the analysis of the results we use an Ising-type spin model in order to elucidate the effects of magnetic interactions on the energetics of the alloy. The rest of the paper is divided into two main sections and conclusions. The theoretical tool is briefly reviewed in Section II and the results are presented and discussed in Section III. 

\section{Method}\label{method}
The calculations are based on the density functional theory\cite{Hohenberg1964,Kohn1965} and performed using the Exact Muffin-Tin Orbitals (EMTO) method.\cite{Vitos2001,Vitos2007} The EMTO method is an improved screened Korringa-Kohn-Rostoker method,\cite{Andersen1994} where the one-electron potential is
represented by large overlapping muffin-tin potential spheres. By
using overlapping spheres, one describes more accurately the crystal
potential, when compared to the conventional non overlapping
muffin-tin approach.\cite{Andersen1998,Vitos2001a,Vitos2000}

The EMTO basis set included $s$,
$p$, $d$, and $f$ orbitals. The one-electron equations were solved
within the scalar-relativistic and soft-core approximation. The generalized gradient approximation in the PBE form was used for the
exchange correlation functional.\cite{Perdew1996} The EMTO Green's function was calculated self-consistently for 32 complex energy points distributed exponentially on a semi-circular contour, which
included states within 1 Ry below the Fermi level. In the one-center
expansion of the full charge density, we adopted an $l$-cutoff of
8 and the total energy was calculated using the full charge-density
technique.\cite{Vitos2001a,Vitos2007} 
For each alloy the calculated equilibrium lattice constant was used. The convergence of the total energy with respect to the number of ${\bf k}$-vectors was tested. It was found that 1240 uniformly distributed ${\bf k}$-vectors within the irreducible wedge of the Brillouin zone was enough for the present purposes. 

The Fe-rich Fe-Cr alloys adopt the body centered cubic (bcc) phase of $\alpha$-Fe. A number of previous works demonstrate that below the magnetic transition temperature (~900-1050 K) the energetics of Fe-Cr alloys with less than ~10\% Cr is well described using the substitutionally disordered ferromagnetic bcc phase.\cite{Olsson2006,Olsson2003} 

Since in the present investigation we map a large concentration region with Cr content approaching zero, the conventional supercell method would require enormously large supercells. Here, we resolve this
difficulty by employing the Coherent Potential Approximation (CPA).\cite{Soven1967} Within the CPA, the alloy components are embedded
in an effective medium, which is constructed in such a way that it
represents, on the average, the scattering properties of the alloy. The EMTO approach in combination with the CPA has been applied successfully in the theoretical study of various structural and electronic properties of alloys and compounds\cite{Vitos2007} demonstrating the accuracy and efficiency needed for the present investigation.

The EMTO results are analyzed by using a spin model\cite{ackland2006} where the magnetic moments obtained from the EMTO calculations are the input data and the output energy is used to calculate the effective chemical potential. This procedure allows a more transparent way to understand the effect of magnetism on the chemistry of the Fe-Cr-X alloys.

\section{Results and discussion}\label{result}
\subsection{Corrosion rate versus balance between bulk and surface chemical potentials}\label{result.a}
Our previous investigations demonstrate that the chemical composition of the close packed surfaces of Fe-Cr alloys follows closely the characteristic threshold-like behavior of the corrosion rate of ferritic stainless steels.\cite{Ropo07,ropo08} In dilute, corrodible, ${\rm Fe}_{1-x}{\rm Cr}_{x}$ alloys ($x{\lesssim}0.05$) the surfaces are exclusively covered by Fe, whereas the Cr-containing surfaces become favorable when the bulk Cr concentration exceeds 10 at.\%, which is close to the generally accepted Cr threshold for corrosion resistant alloys. The discovered threshold point in the concentration of bulk Cr was found to be a consequence of the reversal of the relative magnitudes of the bulk and surface effective chemical potentials ($\mu_{\rm Fe}-\mu_{\rm Cr}$), which in turn reflects the peculiar electronic and magnetic structure of Fe-rich Fe-Cr alloys.\cite{ackland2006,Olsson2003,Olsson2006,Olsson2007,klaver2006,Nonas1998,herper2002a,herper2002b,kiejna2008,ruban2008,ropo11} 

\begin{figure*}[!ht]
\includegraphics[width=0.7\textwidth,angle=-90]{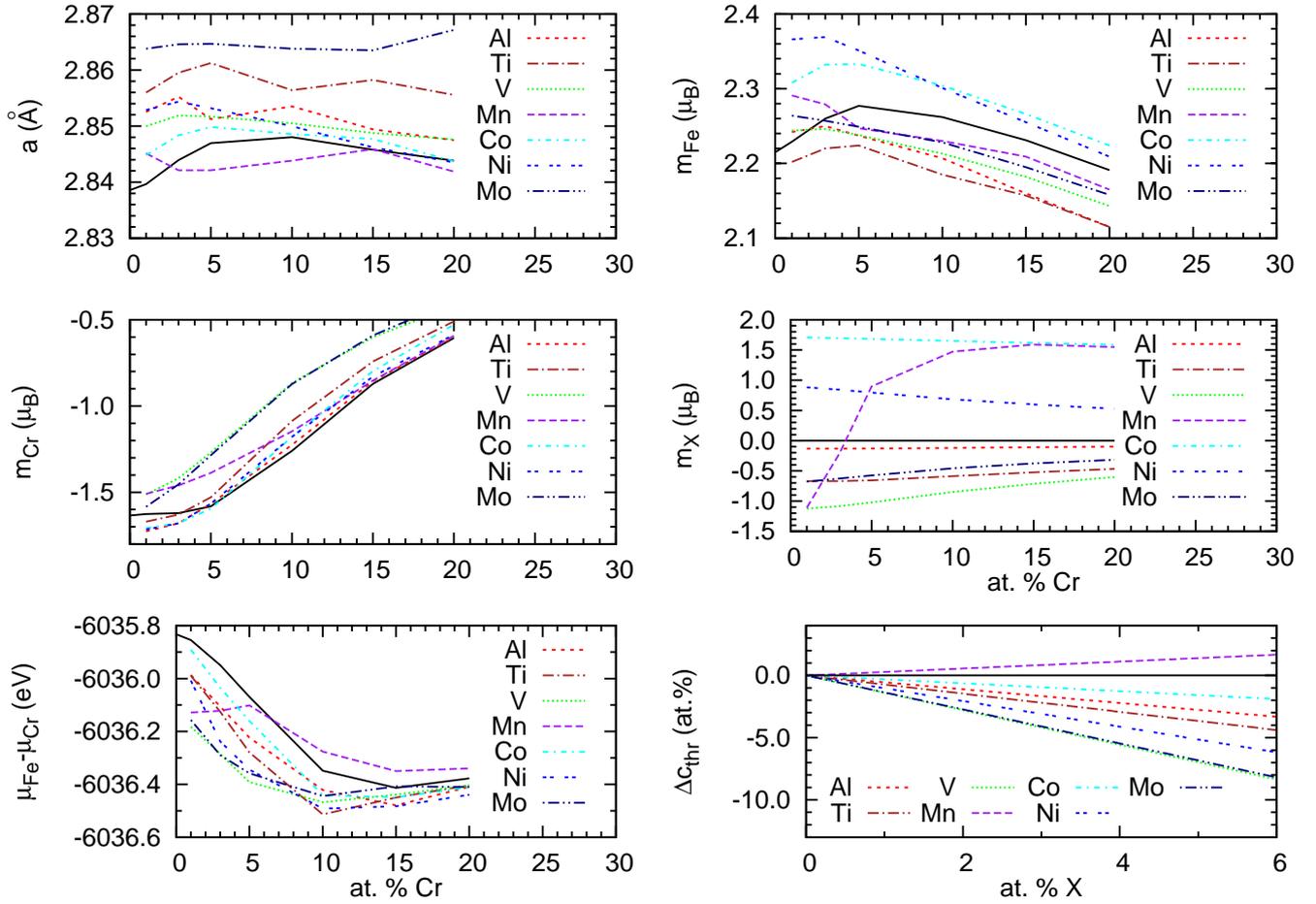}
\caption{(Color online) Results of the EMTO calculations. Lattice parameter ($a$), magnetic moment per atom ($m$, in Bohr magnetons $\mu_{\rm B}$), effective chemical potential in bulk $(\mu_{\rm Fe}-\mu_{\rm Cr})$ and the shift of the Cr threshold $\Delta c_{\rm thr}$ of Fe$_{1-x-0.05}$Cr$_x$X$_{0.05}$ where X is the alloying component (Al, Ti, V, Mn, Co, Ni, Mo). Continuous black curve shows the results for Fe-Cr. $\Delta c_{\rm thr}$ as a function of X content is a linear approximation based on 5 at.\% X data.} \label{figure1}
\end{figure*}

In the present work we extend the previous investigations to questions to what extent the balance of the chemical potentials of Fe and Cr in bulk alloy can be tuned by doping Fe-Cr with typical minor alloying elements commonly used in commercial steels. As a reference for the alloying effect, we use the aforementioned threshold value of bulk Cr which corresponds to the breaking of the pure Fe surfaces. We calculate the effective chemical potential in bulk ($(\mu_{\rm Fe}-\mu_{\rm Cr})^{\rm bulk}$) for Fe-Cr-X alloys containing 5 at.\% X (X=Al, Ti, V, Mn, Co, Ni, Mo) and compare the calculated $(\mu_{\rm Fe}-\mu_{\rm Cr})^{\rm bulk}$ (Fig.~\ref{figure1}) to the corresponding effective chemical potential at the surface of binary Fe-Cr
($(\mu_{\rm Fe}-\mu_{\rm Cr})^{\rm  surface}$), determined in our earlier work.\cite{Ropo07,airiskallio2009,airiskallio2010} 

In alloys for which the surface energy of the third component X is higher than that of Fe, the surfaces are expected to be covered mainly by Fe atoms. Therefore, in these cases the effective chemical potential $\mu_{\rm Fe}-\mu_{\rm Cr}$ at the surface can be approximated by that of the binary Fe-Cr alloy and in these alloys the threshold value of bulk Cr corresponds to the onset of the breaking of the pure Fe surface by Cr atoms. On the other hand, for those alloys where the surface energy of the third component X is lower than that of Fe (Al and Ni, Ref. \onlinecite{vitos1998}), the surfaces of the Fe-Cr-X alloys are expected to be covered mainly by X atoms. In these cases the lowering of the threshold value of bulk Cr corresponds to the increasing of the driving force for moving an Fe atom, compared to that of Cr, from the surfaces to the bulk of the alloy. 

By comparing the effective chemical potential $\mu_{\rm Fe}-\mu_{\rm Cr}$ in bulk and at the surface we are able to estimate, for Fe-Cr-X alloys, the threshold concentrations of bulk Cr, which again are expected to be related to the onset of the corrosion resistance in Fe-Cr based alloys. The first-principles EMTO results for the threshold values ($c_{\rm thr}$) of bulk Cr in Fe$_{1-x-0.05}$Cr$_{x}$X$_{0.05}$ alloys are listed in Table~\ref{t1}.
\begin{table}[ht]
\begin{center}
\begin{tabular}{l|cccccccc}
\hline
\hline
X& Al& Ti& V& Mn& Co& Ni& Mo& Cr\\\hline
$c_{\rm thr}$ (at.\%)& 5& 4& 1& 9& 6& 3& 1& 8\\
\hline
\hline
\end{tabular}
\end{center}
\caption{Theoretical results for the threshold values of bulk Cr corresponding to the reversal of the directions of the Cr and Fe driving forces towards the surface in Fe$_{1-x-0.05}$Cr$_{x}$X$_{0.05}$ alloys. The data is obtained by determining the intersection point of the calculated $(\mu_{\rm Fe}-\mu_{\rm Cr})^{\rm  bulk}$ (Fig.~\ref{figure1})  and $(\mu_{\rm Fe}-\mu_{\rm Cr})^{\rm  surface}$.\cite{Ropo07} The last column (Cr) shows the threshold value for Fe-Cr alloys, which agrees well with the experimentally observed onset point of corrosion protection in Fe-Cr.}\label{t1}
\end{table}
\subsection{Chromium threshold as an indication of the energetics of the Fe-Cr based alloys}\label{result.b}
\begin{figure}[h]
\includegraphics[width=0.32\textwidth,angle=-90]{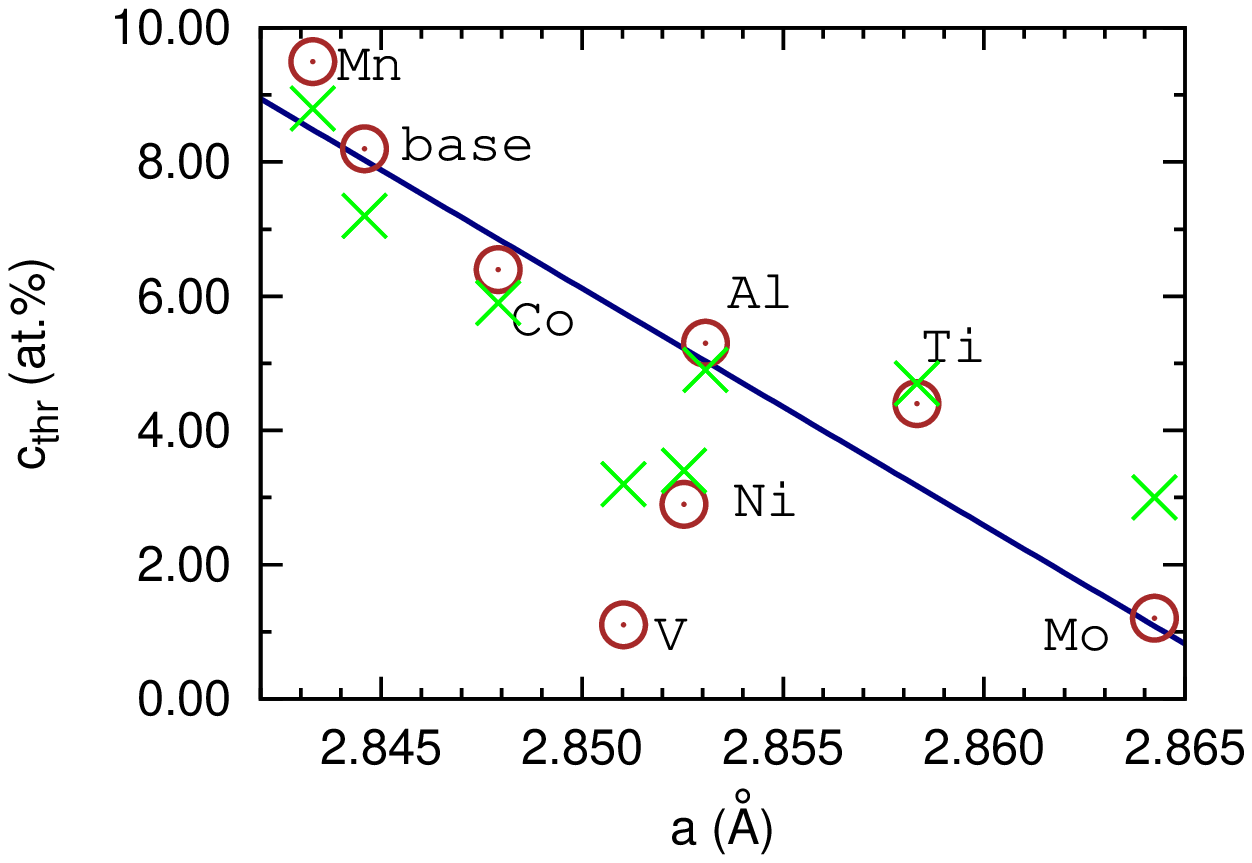}
\caption{(Color online) Threshold of the Cr concentration in Fe$_{1-x-0.05}$Cr$_{x}$X$_{0.05}$ alloys (X = Al, Ti, V, Mn, Co, Ni,  Mo; base refers to Fe-Cr) as a function of the lattice parameter. The brown circles show the EMTO results and the green crosses show the spin-model results. The straight line is a guide for an eye showing the general trend with the volume.} \label{figure.a}
\end{figure}

Plotting the data shown in Table~\ref{t1} as a function of the lattice parameter ($a$) of the alloy (Fig.~\ref{figure.a}) reveals an approximate decreasing trend of the Cr threshold value with the volume of the alloy. The observed general trend shown in the relation between the Cr threshold value and the volume of the alloy can be understood by considering the total energy of the alloy. The approximate connection between the threshold value and the volume can be elucidated as follows:
\begin{itemize}
\item[($i$)]
The increased volume is related to the decreased bonding between the atoms, and consequently to the increased total energy of the alloy.
\item[($ii$)]
The increasing total energy of dilute Fe-Cr alloys decreases the magnitude of the negative slope of the mixing enthalpy of ${\rm Fe}_{1-x}{\rm Cr}_{x}$ at $x\approx 0$. This is due to the characteristic shape of the mixing enthalpy of ${\rm Fe}_{1-x}{\rm Cr}_{x}$ alloys at small $x$.\cite{Ropo07} 
\item[($iii$)]
The relation between the effective chemical potential of bulk Fe-Cr and the slope of the mixing enthalpy ($\Delta$H) of the alloy\cite{ropo06} 
\begin{equation}
(\mu_{\rm Fe}-\mu_{\rm Cr})^{\rm bulk} \approx -\frac{\partial \Delta H}{\partial x} + {\rm constant} \label{eqn.spin}  	
\end{equation}
shows that the effective chemical potential at $x\approx 0$ consequently decreases. 
\item[($iv$)]
Finally, the decrease of $(\mu_{\rm Fe}-\mu_{\rm Cr})^{\rm bulk}$ at small $x$ shifts the threshold value $c_{\rm thr}$ to lower Cr contents. 
\end{itemize} 
At this point it should be noted that vanadium deviates significantly from the general trend of $c_{\rm thr}$ with volume showing the largest reduction in the bulk Cr threshold with increasing volume (Fig.~\ref{figure.a}). Therefore, the threshold value of the bulk Cr as a function of minor alloying element needs to be investigated in more detail. 
\subsection{Establishing the role of magnetic interactions in the thermodynamics of Fe-Cr and its derivatives}\label{result.c}
Because the phase stability of Fe-Cr alloys depends significantly on the magnetic interactions between the constituent atoms,\cite{kissavos2006,ackland2006,ackland2009,Ropo07,ropo08,ropo11} it is instructive to concentrate on the magnetic properties in the present case too. In order to trace the role of the additional alloying elements (X) on the magnetic contribution to the bulk threshold of Cr in Fe-Cr-X alloys we analyze our {\em ab initio} results by using a spin  Hamiltonian
\begin{equation}
H_{ij}=(S_i+S_j)\sigma_i\sigma_j/2+(1-S_iS_j)\sigma_i\sigma_j/2. \label{eqh}
\end{equation}
A similar Hamiltonian was employed by Ackland  (Ref. \onlinecite{ackland2006}) in his study of the Fe-Cr systems. The total energy of the $N$ atom system ($E_{\rm tot}$) per atom then reads
\begin{equation}
E_{\rm tot}/{\rm atom}=\frac{1}{N}\sum_{i=1}^{N}\frac{1}{2}\sum_{j=1}^{14}H_{ij}=\frac{1}{N}\sum_{i=1}^{N}H_i=<H_i>,\label{eqe}
\end{equation}
where the $j$ summation runs over the nearest and next nearest neighbors of the site $i$ (14 sites in the bcc lattice\cite{ackland2006}). Applying equation (\ref{eqe}) to binary   Fe-Cr alloy, we adopt $S_{\rm Fe}=-1$, $S_{\rm Cr}=1$, and  $\sigma_{\rm Fe}$ and $\sigma_{\rm Cr}$ are the magnetic moments of Fe and Cr atoms, respectively. This yields ferromagnetic Fe-Fe coupling and antiferromagnetic coupling of Cr to its neighbors. Within the CPA formalism, we obtain for the Fe$_{1-x}$Cr$_x$ alloy
\begin{eqnarray}
<H_i>&=&x<H_{\rm Cr}>+(1-x)<H_{\rm Fe}>\nonumber \\
&=&7(x^{2}\sigma^{2}_{\rm Cr}+2x(1-x)\sigma_{\rm Cr}\sigma_{\rm Fe}\nonumber \\
&-&(1-x)^{2}\sigma^{2}_{\rm Fe}).\label{eqn3}
\end{eqnarray}
\begin{figure}[h]
\includegraphics[width=0.4\textwidth,angle=-90]{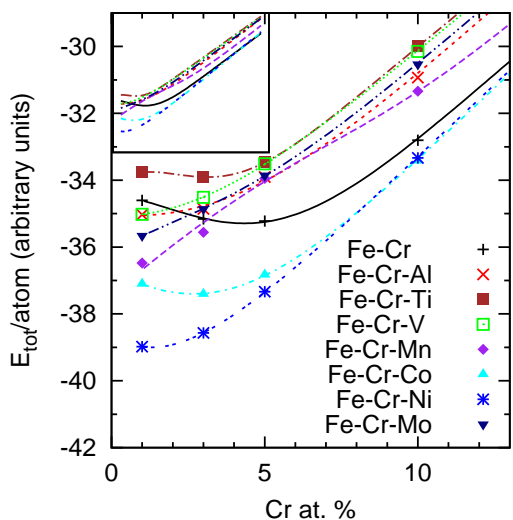}
\caption{(Color online) The total energy per atom obtained from the spin model. The symbols refer to the results of the spin model (Eqn. (\ref{eqn3})) and curves are functions fitted to the spin model data. The inset shows the same results but from 1 to 20 at.\% Cr emphasizing the almost linear behavior of the energy as obtained from the spin model within the 10--20 at.\% Cr range.} \label{figure.b}
\end{figure}
Since we are interested in the relative energy balance of Cr in bulk, we extend the above spin model to Fe-Cr-X by merging the X atoms with the Fe atoms to form an effective Fe matrix into which Cr atoms are dissolved. For individual spins $\sigma$ we use the calculated magnetic moments (Fig.~\ref{figure1}). 
The spin of the effective Fe matrix is taken to be the average of the magnetic moments of the Fe and X atoms weighted by their concentrations. As Fig.~\ref{figure.b} shows, the alloying of Fe-Cr changes both the curvature and the position of the minimum of the total energy (mixing enthalpy), particularly around the point of 5  at.\% Cr which is an important region with respect to the bulk Cr threshold. Because the bulk effective chemical potential $(\mu_{\rm Fe}-\mu_{\rm Cr})^{\rm bulk}$ is a key quantity in determining the bulk Cr threshold and since the effective potential is approximately the derivative of the total energy (Eqn.~(\ref{eqn.spin})) we calculate the derivatives of the spin model total energies from Fig.~\ref{figure.b}. The result is shown in Fig.~\ref{figure.c}.
\begin{figure}[h]
\includegraphics[width=0.4\textwidth,angle=-90]{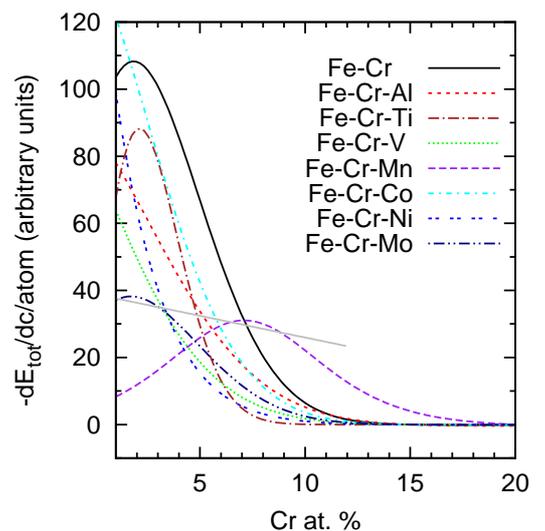}
\caption{(Color online) The derivative of the spin model total energy per atom (spin model, Fig.~\ref{figure.b}) multiplied by $-1$ and shifted to the common zero value at Cr at.\% = 20. The short gray line segment shows the level of the EMTO surface effective chemical potential ($(\mu_{\rm Fe}-\mu_{\rm Cr})^{\rm surface}$, from Ref.~\onlinecite{Ropo07}) relative to the Fe-Cr curve.} \label{figure.c}
\end{figure}
The derivatives of the total energy of the spin model have approximately the same shape as the EMTO effective chemical potentials $(\mu_{\rm Fe}-\mu_{\rm Cr})^{\rm bulk}$, shown in  Fig.~\ref{figure1}. Plotting in Fig.~\ref{figure.c} the surface chemical potential $(\mu_{\rm Fe}-\mu_{\rm Cr})^{\rm surface}$ of Fe-Cr alloy, we can estimate the bulk Cr thresholds ${\Delta}c_{\rm thr}$ by determining the positions of the intersections of the $(\mu_{\rm Fe}-\mu_{\rm Cr})^{\rm bulk}$ and $(\mu_{\rm Fe}-\mu_{\rm Cr})^{\rm surface}$ curves. These estimated critical compositions are shown by green crosses in Fig.~\ref{figure.a}. The spin model based Cr threshold deviates significantly from that of the EMTO calculations in the case of V and Mo. What comes to the qualitative agreement, it is of course expected that the spin model cannot compete with the {\em ab initio} calculations, but what is more important here is that the general trend predicted by the spin model agrees with the EMTO result. This suggests that the spin model is capable of capturing the main features of the Fe-Cr-X alloy and to reveal the underlying magnetic interactions. The obtained shifts of the thresholds of bulk Cr (${\Delta}c_{\rm thr}$) due to the alloying in the spin model agree surprisingly well with those obtained directly by the EMTO calculation (Fig.~\ref{figure.bar}).
\begin{figure}[h]
\includegraphics[width=0.32\textwidth,angle=-90]{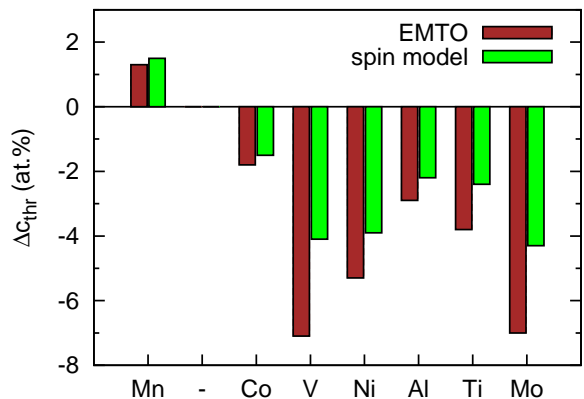}
\caption{(Color online) The shifts of the bulk Cr threshold ${\Delta}c_{\rm thr}$ from that of the Fe-Cr case (EMTO result: 8 at.\%, spin model result: 7 at.\%) due to 5 \% alloying with Al, Ti, V, Mn, Co, Ni, and Mo, obtained by the EMTO method (brown bars) and the spin model (green bars).} \label{figure.bar}
\end{figure}
\begin{figure}[h]
\includegraphics[width=0.35\textwidth,angle=-90]{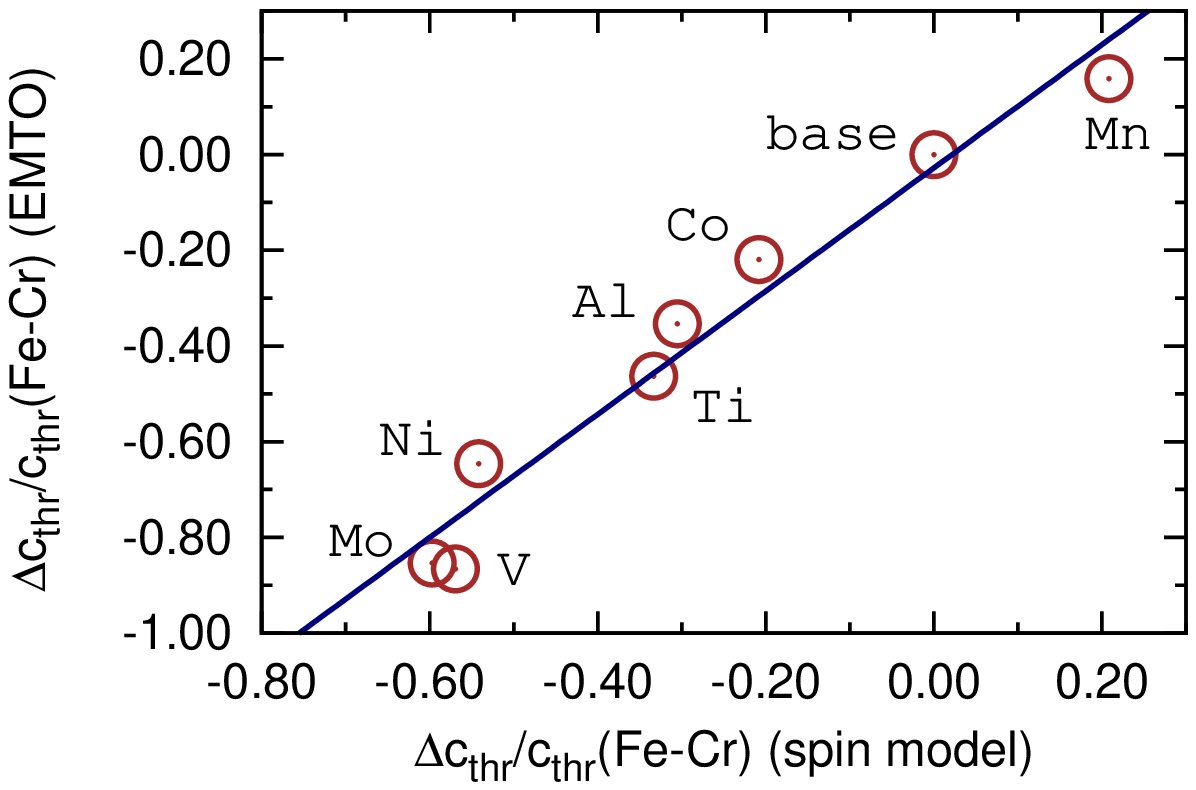}
\caption{(Color online) Relative shift of the threshold Cr concentration ${\Delta}c_{\rm thr}/c_{\rm thr}{\rm (Fe-Cr)}$ in Fe$_{1-x-0.05}$Cr$_{x}$X$_{0.05}$ with (X = Al, Ti, V, Mn, Co, Ni, and Mo; base refers to Fe-Cr). The line shows the linear fit to the data $f(x)=a+bx$, $a = -0.028$ $b = 1.29$.} \label{figure.aa}
\end{figure}
Presenting the results using relative quantities, 
\begin{equation}
\frac{ {\Delta}c_{\rm thr} }{ c_{\rm thr}{\rm (Fe-Cr)} }
=
\frac{ c_{\rm thr}{\rm (Fe-Cr-X)}-c_{\rm thr}{\rm (Fe-Cr)} }{  c_{\rm thr}{\rm (Fe-Cr)} } 
\end{equation}
shown in Fig.~\ref{figure.aa}, it is realized that there is a significant correlation between the relative thresholds of the EMTO calculation and the spin model estimate demonstrating definitely that the effective chemical potential of Fe and Cr in the investigated alloys is to a large extent determined by magnetic interactions.
We close our discussion by analyzing the EMTO results shown in Fig.~\ref{figure1}. The general trend seen in Fig.~\ref{figure1} is that all investigated alloying components lower the $(\mu_{\rm Fe}-\mu_{\rm Cr})^{\rm bulk}$ for low-Cr alloys. When Cr content increases, this trend is diminished or reversed (Mn). When Cr content reaches 15--20 at.\% the effect of the additional alloying is small. Manganese shows the most peculiar behavior: below 5 at.\% Cr, Mn strongly decreases $(\mu_{\rm Fe}-\mu_{\rm Cr})^{\rm bulk}$ whereas above 10 at.\% Cr it increases $(\mu_{\rm Fe}-\mu_{\rm Cr})^{\rm bulk}$.

Maybe the most interesting data in Fig.~\ref{figure1} is related to the magnetic moments of Mn. It provides an illustrative evidence for the decisive role of the magnetic moments of the alloying elements in the relative balance between the bulk chemical potentials of Fe and Cr in Fe-Cr-X alloys. While Cr concentration changes from 0 to 5 at.\% the Mn moment changes approximately from $-1$ to $+1$ Bohr magneton ($\mu_{\rm B}$) i.e.\ practically through the whole range of the moments of the investigated additional alloying elements (from V to Co). Correspondingly the  $(\mu_{\rm Fe}-\mu_{\rm Cr})^{\rm bulk}$ changes from the value of the Fe-Cr-V case to that of the Fe-Cr-Co case revealing the crucial role of the magnetic moment of the alloying element on the chemical potentials in Fe-Cr alloys.
\section{Conclusions}
Using density functional theory, implemented in the EMTO formalism, and a spin model, we investigated the effective chemical potential $(\mu_{\rm Fe}-\mu_{\rm Cr})^{\rm bulk}$ and showed that the magnetic moment of the doping atom plays a significant role in the energy balance of Fe and Cr atoms in Fe-Cr-X alloys. The employed computational method provides an efficient tool to scan the effects of different alloying components at any concentration level. Since the stainless property of Fe-Cr based alloys is related to the presence of Cr near the alloy surface, our findings are useful in understanding and explaining the corrosion resistance of the Fe-rich Fe-Cr based alloys. Using the obtained results for the bulk it is possible to estimate the effect of a third component on the relative content of Fe and Cr at the surface.

\section{Acknowledgements}

The computer resources of the Finnish IT Center for Science (CSC) and Mgrid project are acknowledged. Financial support from the Academy of Finland (Grant No.\ 116317) (EA, EN, KK) and Outokumpu Foundation (EA) are acknowledged. The Swedish Research Council, the European Research Council, the Swedish Steel Producer's Association (LV, BJ), the Hungarian Research Fund (OTKA project 84078) (LV) and the G\"oran Gustafson Foundation (MP) are also acknowledged.

\end{document}